\documentclass[showpacs,aps,twocolumn,pre,amsmath,amssymb]{revtex4}

\usepackage{epsfig}

\begin{document}



\title{Short-time critical dynamics
at perfect and non-perfect surface}

\author{Shizeng Lin$^{1,2}$ and Bo Zheng$^{1}$}

\affiliation{$^1$ Zhejiang University, Zhejiang Institute of Modern
Physics, Hangzhou 310027, P.R. China\\
$^2$ Computational Materials Science Center, National Institute for
Materials Science, Sengen 1-2-1, Tsukuba 305-0047, Japan\\}

\begin{abstract}
We report Monte Carlo simulations of critical dynamics far from
equilibrium on a perfect and non-perfect surface in the $3d$ Ising
model. For an ordered initial state, the dynamic relaxation of the
surface magnetization, the line magnetization of the defect line,
and the corresponding susceptibilities and appropriate cumulant is
carefully examined at the ordinary, special and surface phase
transitions. The universal dynamic scaling behavior including a
dynamic crossover scaling form is identified. The exponent $\beta_1$
of the surface magnetization and $\beta_2$ of the line magnetization
are extracted. The impact of the defect line on the surface
universality classes is investigated.
\end{abstract}


\maketitle

\section{Introduction}
The breakdown of space and time translation invariance leads to
geometric and temporal surface effects. The former is very common in
a system whose spatial correlation length is comparable to its
dimensions. Such effects become even more important when nano-scale
materials are concerned. In a recent experiment, for example, an
anomalous temperature profile of the phase transitions was observed
in the presence of a ferromagnetic surface \cite{tor05}. The latter
occurs in a nonequilibrium system, which is prepared by suddenly
quenching the system to its critical temperature from any given
initial condition.

The breakdown of space translation invariance modifies the critical
behaviors near geometric surface and new critical exponents must be
introduced. There may exist several universality classes in one bulk
system in the presence of free geometric surface. The critical
behavior of geometric surface has been extensively studied, and the
equilibrium phase diagram has been well established in the past
decades \cite{bin87,die87,die97,ple04}. However, most previous
studies concentrated on the static behaviors
\cite{bin72,bin84,lan90,rug95,den05} and the dynamics in the
long-time regime\cite{die83,kik85,die94}, i.e. system only with
geometric surface. The critical dynamics of surface in the {\it
macroscopic} short-time regime, i.e., when the system is still far
from equilibrium, is much less touched \cite{ple04}.

On the other hand, for a system quenched to its critical
temperature, because there is no characteristic time scale, the
temporal surface has long-lasting effect. This effect has very
important consequences. One is that in nonequilibrium dynamic
relaxation of magnetization, if in the initial state, there is
small, nonvanishing magnetization $m_0\ll 1$, the magnetization
grows as $m_0 t^\theta$ with $\theta$ being a new nonequilibrium
dynamic exponent\cite{jan89}. In such short-time critical dynamics,
there exist two competing nonequilibrium dynamic processes. One is
the domain growth with scaling dimension $x_i$ and the other is the
critical thermal fluctuation with scaling dimension $x=\beta/\nu$.
Because the spatial correlation length $\xi$ grows as $t^{1/z}$, we
can relate $\theta$ to $x_i$ and $x$ by $\theta=\frac{x_i-x}{z}$.
Generally $x_i$ is larger than $x$ and the net effect is the domain
growth in the nonequilibrium relaxation process. This short-time
critical dynamics of bulk has been established in the past decade,
and successfully applied to different physical systems
\cite{jan89,hus89,zhe98,zhe99,god02,cal05}. Based on the short-time
dynamic scaling, new techniques for the measurements of both dynamic
and static critical exponents as well as the critical temperature
have been developed \cite{li95,luo98,sch00}. Recent progresses can
be found partially in Refs. \cite{zhe98,zhe03a,yin04,yin05,fed06}.

Obviously, the physical phenomena are more complicated, when both
temporal and geometric surfaces are considered. The interplay
between both surfaces embraces many interesting physics and is worth
for careful studies\cite{ple04,rit95}. Recently it is reported that
in non-equilibrium states, the surface cluster dissolution may take
place instead of the domain growth \cite{ple04b,ple05}. In these
studies, the dynamic relaxation starting from a high-temperature
state is concerned.

The impact of defect on geometric surface is also of great concern.
The presence of imperfection may alter the surface university
classes and even the phase diagram. The former is easily signaled
from the non-equilibrium dynamics as in the case of
bulk\cite{yin04,yin05,luo01}.

In this paper, we study the short-time critical dynamics on {\it a
perfect and non-perfect surface} with Monte Carlo simulations. We
generalize the universal dynamic scaling behavior to the dynamic
relaxation at geometric surfaces, starting from the {\it ordered}
state. At the ordinary, special and surface phase transitions, the
dynamic scaling behavior of the surface magnetization,
susceptibility and appropriate cumulant are identified. The static
exponent $\beta_1$ of the surface magnetization and $\beta_2$ of the
line magnetization of the defect line are extracted from the dynamic
behavior in the macroscopic short-time regime. The robustness of
surface university class against extended defect is investigated by
means of non-equilibrium dynamics. The surface transition and
special transition can also be detected from the short-time
dynamics.

The remaining part of this paper is organized as follows. In Sec.
\ref{sec2}, the definition of the model and the short-time dynamic
scaling analysis are presented. In Sec. \ref{sec3} and \ref{sec4},
the dynamic relaxation on a perfect and non-perfect surface is
studied. In Sec. \ref{sec5}, the results are summarized.

\section{Model and dynamic scaling analysis}\label{sec2}

\subsection{Model}

The Hamiltonian of the $3d$ Ising model with Glauber dynamics and
line defect on free surface in the absence of external magnetic
field can be written as the sum of bulk interactions, surface
interactions and line interactions,
\begin{equation}
\begin{array}{l}
H=-J_b\sum_{<xyz>}^{bulk}\sigma_{xyz}
\sigma_{x'y'z'}-J_s\sum_{<xy>}^{surface}\sigma_{xyz} \sigma_{x'y'z'}
\\
\\
- J_{l}\sum_{<y>}^{defect}\sigma_{xyz} \sigma_{x'y'z'},
\end{array}
\end{equation}
where spin $\sigma$ can take values $\pm 1$ and $<xyz>$ indicates
the summation over all nearest neighbors. The first sum runs over
all links including at least one site that does not belong to the
surface, whereas the second sum runs over all surface links
excluding the links that both sites are inside the defect line. The
last summation extends over all links which belong to the defect
line. $J_b$, $J_s$ and $J_l$ are the coupling constants for the
bulk, surface and defect line respectively. For ferromagnetic
materials, $J_b$ and $J_s$ are positive. It is generally believed
that the dynamic universality class of Glauber dynamics is
insensitive to the detailed algorithm used as long as the updating
algorithm is local. Here we use Metropolis spin-flip algorithm.
Without explicitly specified, the dynamic exponent refers to the
Ising model with Glauber dynamics in the following discussions.

\begin{figure}[b]
\psfig{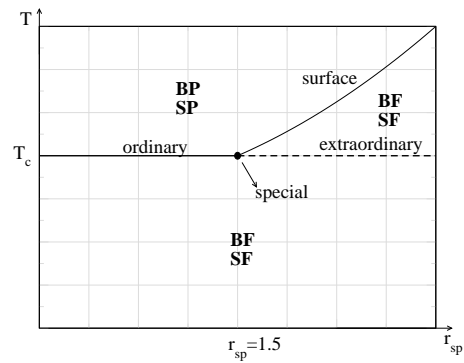}
\caption{\label{f0}Schematic phase diagram for the semi-infinite
Ising model with bulk coupling $J_b$ and surface coupling ratio
$J_s/J_b$. $T_b$ is the bulk transition temperature and the
ferromagnetic bulk is denoted by FB while the paramagnetic is
denoted by PB. The surface phases are labeled FS for a ferromagnet
and PS for a paramagnet.}
\end{figure}

For a perfect surface, i.e., $J_l=J_s$, it is well known that there
exists a special threshold $r_{sp}\equiv J_s/J_b$ in equilibrium.
For $J_s/J_b<r_{sp}$, the surface undergoes a phase transition at
the bulk transition temperature $T_b$, due to the divergent
correlation length in the bulk. This phase transition is called {\it
the ordinary transition}, and the critical behavior is independent
of $J_s/J_b$. See Fig. \ref{f0}. This is a {\it strong}
universality. For $J_s/J_b>r_{sp}$, the surface first becomes
ferromagnetic at a surface transition temperature $T_s>T_b$, while
the bulk remains to be paramagnetic. If the temperature is further
reduced, the bulk becomes also ferromagnetic at $T_b$. The former
phase transition is called {\it the surface transition} and the
latter is called {\it the extraordinary transition}. It is generally
believed that the surface transition belongs to the universality
class of the $2d$ Ising model \cite{bin87,die87}. Around $r_{sp}$
occurs the crossover behavior. At exactly $J_{s}/J_b=r_{sp}$, the
lines of surface transition, ordinary transition and extraordinary
transition meet at this multicritical point with new surface
exponents. The surface and bulk become critical simultaneously at
this point and this phase transition is called {\it the special
transition}. The best estimate of $r_{sp}$ for the $3d$ Ising in
equilibrium is $1.5004(20)$ \cite{rug92}.

For a non-perfect surface, we introduce a defect line with coupling
strength $J_l$ onto the surface. Generally speaking, the impact of
imperfection on a surface is two fold. Take a surface with random
bond disorder as an example. The randomness may reduce the surface
transition temperature, and alter the global phase diagram of a
semi-infinite system. For example, the special transition point of
the Ising model with a amorphous surface is located at
$r_{sp}=1.70(1)$ \cite{ple98b}, noticeably larger than that of the
Ising model with a perfect surface $r_{sp}=1.5004$. Another effect
of the randomness is that it may change the universality class of
the surface. The relevance or irrelevance of random imperfections on
the pure surface can be assessed by the Harris-type criterion
\cite{die90}. The extended-Harris criterion states that for a
surface with random bond disorder, the disorder is relevant for
$\alpha_{11}>0$ but irrelevant for $\alpha_{11}<0$. Based on this
criterion, the random surface coupling of the Ising model is
irrelevant at the ordinary transition since $\alpha_{11}<0$. In this
case, it was rigorously proved $\beta_1^{dis}=\beta_1^{ord}$ by
Diehl based on the Griffiths-Kelly-Sherman inequality, where
$\beta_1^{dis}$ is the critical exponent at the ordinary transition
on a random bond surface\cite{die98a}. The situation is less clear
at the special transition, for $\alpha_{11}$ is very close to $0$.
Recent simulations suggest that $\alpha_{11}<0$ and hence the
disorder is irrelevant\cite{den05}. The irrelevance at the special
transition has also been reported in Ref. \cite{ple98b}. At the
surface transition, the surface is equivalent to the $2d$ Ising
model. The disorder only leads to logarithm correction (see
\cite{luo01} and reference therein). In the case of defect line, the
defect doesn't shift the transition temperatures of the surface
transition, and therefore, the special transition point $r_{sp}$ at
which the surface transition line and ordinary transition line meet
remains to be unchanged \cite{fis67}. We only consider the
robustness of the ordinary, special and surface transition in the
presence of line defect.
\subsection{Dynamic scaling analysis}

\begin{table}[h]\centering
\caption{The bulk critical temperature and critical exponents of the
$3d$ Ising model.}
\begin{tabular}{ c | c | c }\hline\hline
$T_c$ & $\nu_{3d}$ & $z_{3d}$\\
 \hline
$4.5115248(6)$ \cite{den03} &  $0.6298(5)$ \cite{fer91} & $2.042(6)$ \cite{jas99}\\
\hline \hline
\end{tabular}
\label{tbl1}
\end{table}

For a dynamic system, which is initially in a high-temperature
state, suddenly quenched to the critical temperature, and then
released to the dynamic evolution of model A, one expects that there
exist universal scaling behaviors already in the {\it macroscopic
short-time regime} \cite{jan89}. This has been shown both
theoretically and numerically in a variety of statistical systems
\cite{jan89,hus89,zhe98,zhe99,god02,cal05,zhe03a,yin05}, and it
explains also the spin glass dynamics. Furthermore, the short-time
dynamic scaling behavior has been extended to the dynamic relaxation
with an ordered initial state or arbitrary initial state, based on
numerical simulations \cite{zhe98,zhe03a,yin05,zhe96,jas99}. Recent
renormalization group calculations also support the short-time
dynamic scaling form for the ordered initial state \cite{fed06}.

On the other hand, Ritschel and Czerner have generalized the
short-time critical dynamics in a homogenous system to that in an
inhomogeneous one, i.e., the systems with a free surface, and
derived the scaling behavior of the magnetization close to the
surface for the dynamic relaxation with a high-temperature initial
state \cite{rit95}. Recent development can be found in Refs.
\cite{ple04b,ple05}. In this paper, we alternatively focus on the
dynamic relaxation with the ordered initial state, and with a
non-perfect surface. As pointed out in the literatures
\cite{zhe98,zhe03a}, the fluctuation is less severe in this case. It
helps to obtain a more accurate estimate of the critical exponents
at surface. From theoretical point of view, it is also interesting
to study the dynamic relaxation with the ordered or even arbitrary
initial state.

Similar to the scaling analysis in bulk
\cite{jan89,zhe98,jas99,fed06}, we phenomenologically assume that,
for dynamic relaxation with ordered initial state, the surface
magnetization decays by a power law,
\begin{equation}
<m_1(t)>\sim t^{-\beta_1/\nu_s z_s}, \label{e30}
\end{equation}
after a microscopic time scale $t_{mic}$. Here $<\cdot>$ represents
the statistical average, $\beta_1$ is the static exponent of the
surface magnetization, $\nu_s$ is the static exponent of the spatial
correlation length, and $z_s$ is the dynamic exponent. This
assumption can be understood by noting that, for nonequilibrium
preparation with ordered initial state $m_0=1$, the dynamic
relaxation is governed by critical thermal fluctuation with scaling
dimension $x_1=\beta_1/\nu_s$. For the {\it ordinary} and {\it
special} transitions where the criticality of surface originates
from the divergence of the correlation length in bulk, there are no
genuine new surface dynamic exponent $z_s$ and static exponent
$\nu_s$. $\nu_s$ and $z_s$ are just the same as those in the bulk,
i.e. $\nu_s=\nu_{3d}$ and $z_s=z_{3d}$, while $\beta_1$ is neither
that of the $2d$ Ising model nor that of the $3d$ Ising model
\cite{die83}. For the {\it surface} transition where the critical
fluctuation of surface is of the universality class of the $2d$
Ising model, it is generally believed that all static and dynamic
exponents are the same as those of the $2d$ Ising model
\cite{bin87,die87}. i.e. $\beta_1=\beta_{2d}=1/8$,
$\nu_s=\nu_{2d}=1$ and $z_s=z_{2d}\approx 2.16(2)$ \cite{zhe98}.

\begin{table*}[t]\centering
\caption{Summary of the surface critical exponents at the ordinary
and special transition in the $3d$ Ising model, as obtained by
different techniques. MF: mean-field, MC: Monte Carlo simulations,
FT: field-theoretical methods, CI: conformal invariance. The data
marked with $*$ are calculated by using scaling law
$2\beta_1+\gamma_{11}=(d-1)\nu_s$.}
\begin{tabular}{c | c | c | c | c | c | c | c | c | c}\hline\hline
    &  MF \cite{bin87} & MC \cite{lan90} & MC \cite{rug95} & MC \cite{ple98} & MC \cite{den05} & MC \cite{rug93}& MC+CI \cite{den03} & FT \cite{die98}& \emph{this work}\\
 \hline
$\beta_1^{ord}$ & $1$ & $0.78(2)$ & $0.807(4)$ & $0.80(1)$ & $0.796(1)$ & $-$ & $0.798(5)$ & $0.796$ & $0.795(6)$\\
$\beta_{1}^{sp}$ & $1/2$ & $0.18(2)$ & $0.238(2)$ & $-$ & $0.229(1)$ & $0.237(5)$ & $-$& $0.263$ &$0.220(3)$\\
$\gamma_{11}^{sp}$ & $1/2$ & $0.96(9)$ & $0.788(1)$ & $-$ & $0.802(3)^*$ & $0.785(11)^*$ & $-$ & $0.734$ &$0.823(4)$\\
 \hline \hline
\end{tabular}
\label{tbl2}
\end{table*}

Another important observable is the second moment of the surface
magnetization, or the so-called time-dependent surface
susceptibility, defined as
\begin{equation}
\chi_{11}=L^2[<m_1^2>-<m_1>^2].
\end{equation}
Simple finite-size scaling analysis \cite{zhe98} reveals that
\begin{equation}
\chi_{11}(t)\sim t^{\gamma_{11}/\nu_s z_s}. \label{e50}
\end{equation}
Here the exponent $\gamma_{11}/\nu_s$ is related to $\beta_1/\nu_s$
by $\gamma_{11}/\nu_s=d-1-2\beta_1/\nu_s$, with $d=3$ being the
spatial dimension of bulk. This is nothing but the scaling law in
equilibrium between the exponent of the surface susceptibility and
the exponent of the surface magnetization. One can also understand
the scaling behavior in Eq.~(\ref{e50}) in an intuitive way. In
equilibrium, $\chi_{11}$ behaves as $\chi_{11} \sim
L^{\gamma_{11}/\nu_s}$ with $L$ being the lattice size. In the
dynamic evolution, $\chi_{11}(t)$ should be related to the
non-equilibrium spatial correlation length $\xi(t)$ with
$\chi_{11}(t) \sim \xi(t)^{\gamma_{11}/\nu_s}$, since the finite
size effect is negligible. Then the growth law $\xi(t) \sim
t^{1/z_s}$ of the non-equilibrium spatial correlation length
immediately leads to Eq.~(\ref{e50}).

Alternatively, one can also construct the appropriate time-dependent
cumulant $U(t)=<m_1^2>/<m_1>^2-1 $. Obviously, $U(t) \sim
t^{(\gamma_{11}+2\beta_1)/\nu_s z_s}$. From the scaling law
$\gamma_{11}/\nu_s=d-1-2\beta_1/\nu_s$, one derives
$(\gamma_{11}+2\beta_1)/\nu_s=d-1$. The scaling behavior of $U(t)$
then reduces to the standard form \cite{zhe98,jas99},
\begin{equation}
U(t) \sim t^{(d-1)/z_s}, \label{e60}
\end{equation}
with $d-1$ being the spatial dimension of the surface.

In other words, from Eqs.~(\ref{e30}) and (\ref{e50}), or from
Eqs.~(\ref{e30}) and (\ref{e60}), we obtain independent measurements
of two critical exponents, e.g., $\beta_1/\nu_s$ and $z_s$.
Alternatively, if we take $\nu_s$ and $z_s$ as input, we have two
independent estimates of the static exponent $\beta_1$ of the
surface magnetization. This may testify the consistency of our
dynamic scaling analysis.

All foregoing equations involve the bulk exponents $\nu_s$ and
$z_s$. Therefor an accurate estimate of the surface critical
exponents $\beta_1$ needs precise values of $\nu_{3d}$ and $z_{3d}$,
as well as $z_{2d}$. Since the $3d$ bulk Ising model has been
extensively studied with various methods, many accurate results of
the critical exponents and transition temperature are available. We
concentrate our attention to the surface exponents and take the bulk
exponents as input. The results of the bulk exponents of the $3d$
Ising model are summarized in Table \ref{tbl1}. The criteria to
choose those values are their relative accuracy, as well as the
methods used to extract these exponents.

\subsection{Simulations}
In this paper, with Monte Carlo simulations we study the dynamic
relaxation of the $3d$ Ising model on a perfect and non-perfect
surface at the transition temperature, quenched from a completely
ordered initial state. The standard Metropolis algorithm is adopted
in the simulations. In order to investigate the surface critical
behavior, we apply the periodic boundary condition in the $xy$ plane
and open boundary condition in the z direction to the $L\times
L\times L$ cubic lattice.

The main results are obtained with the lattice size $L=128$ and
$L=80$, and additional simulations with other lattice sizes are also
performed to study the finite-size effect. For a perfect surface,
the surface magnetization is defined as
\begin{equation}
m_1=\frac{1}{2L^2}\sum_{xy}^L(\sigma_{xy1}+\sigma_{xyL}),
\end{equation}
and its critical exponent is denoted by $\beta_1$. For a non-perfect
surface, the defect line is placed at surface position $x=L/2$ and
the line magnetization is defined as
\begin{equation}
m_2=\frac{1}{2L}\sum_{xy}^L(\sigma_{\frac{L}{2}y1}+\sigma_{\frac{L}{2}yL}),
\end{equation}
and its critical exponent is denoted by $\beta_2$. The spin
$\sigma_{xyz}$ denotes the spin sitting at site $(x,y,z)$. We
measure the surface and line magnetization during the nonequilibrium
relaxation. We average from $5000$ to $20 000$ runs with different
random numbers to achieve a good statistics. Error bars are
estimated by dividing the total samples into two subgroups, and by
measuring the exponents at different time intervals. Most of the
simulations are carried out on the Dawning 4000A supercomputer. The
total CPU time is about $3$ node-year.

\section{Short-time dynamics on a perfect surface}\label{sec3}

In this section we study the nonequilibrium critical dynamics on a
perfect surface, i.e. $J_l=J_s$. To investigate the critical
behavior on the surface, it is important to know the special
transition point $r_{sp}$. For a perfect surface of the $3d$ Ising
model with ferromagnetic interactions, there exist rather accurate
estimates of $r_{sp}$ in equilibrium, e.g., $r_{sp}=1.5004(20)$ in
Ref.~\cite{rug92}. We adopt this value as the special transition
point. As illustrated later, the special transition point $r_{sp}$
can also be estimated from the scaling plot of a dynamic crossover
scaling relation.

\begin{figure}[b]
\psfig{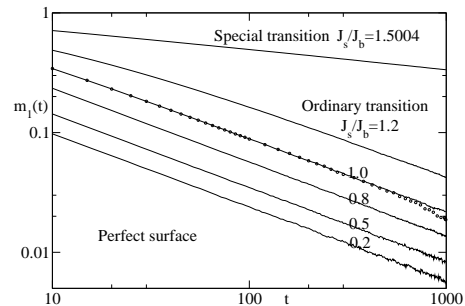} \caption{\label{f1a}Dynamic
relaxation of the surface magnetization is displayed with solid
lines on a double-log scale, at the ordinary transition with various
$J_s/J_b$, and at the special transition $J_s/J_b=r_{sp}=1.5004$.
The temperature is set to the bulk critical temperature $T_c$, and
the lattice size is $L=80$. Open circle are the data for
$J_s/J_b=1.0$ and $L=40$. Well away from the special transition, the
slope of the curves is independent of $J_s/J_b$.}
\end{figure}

\begin{figure}[t]
\psfig{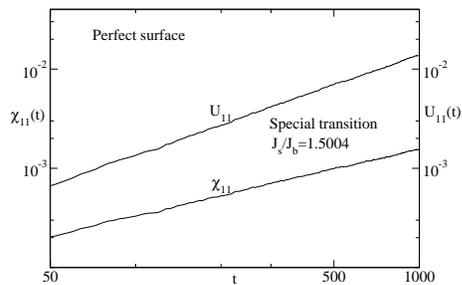}
\caption{\label{f1b}Dynamic relaxation of the surface susceptibility
and cumulant at the special transition is plotted on a double-log
scale. The lattice size is $L=128$ and $T=T_c$.}
\end{figure}

For the ordinary phase transition, the dynamic relaxation of the
surface magnetization with different $J_s/J_b$ are shown in Fig.
\ref{f1a}. The curves of $J_s/J_b=1.0$ with $L=40$ and $L=80$
overlap up to $t \ge 300$MCS(Monte Carlo sweep per site). It
confirms that the finite-size effect is negligibly small for $L=80$
up to at least $t=1000$MCS, since the correlating time of a finite
system increases by $t_L \sim L^z$. In Fig. \ref{f1a}, a power-law
behavior is observed for all $J_s/J_b$. The microscopic time scale
$t_{mic}$, after which the short-time universal scaling behavior
emerges, in other words, after which the correction to scaling is
negligible, gradually increases as the surface coupling is being
enhanced. For $J_s/J_b=0.2$, $t_{mic}\sim 10$MCS, while for
$J_s/J_b=1.2$, $t_{mic} \sim 100$MCS.

A direct observation in Fig. \ref{f1a} is that the curves for
$J_s/J_b < r_{sp}$ are parallel to each other. By fitting these
curves to Eq.~(\ref{e30}), we obtain $\beta_1^{ord}=0.790(7)$,
$0.792(6)$, $0.795(6)$, $0.786(6)$ and $0.755(12)$ for
$J_s/J_b=0.2$, $0.5$, $0.8$, $1.0$ and $1.2$ respectively. The
values of $\beta_1^{ord}$ at $J_s/J_b=0.2$, $J_s/J_b=0.5$ and
$J_s/J_b=0.8$ are well consistent with each other within error. It
indicates that the ordinary transition is universal over a wide
range in the $J_s/J_b$ space. Deviation occurs for $J_s/J_b > 1.0$
and manifests itself as the effect of the crossover to the special
transition. This is in agreement with the observation in
Ref.~\cite{lan90}. From our analysis, $\beta_1^{ord}=0.795(6)$ is a
good estimate for the ordinary transition. In Table \ref{tbl2}, we
have compile all the existing results which were obtained with
simulations and analytical calculations in equilibrium, and our
measurements from the non-equilibrium dynamic relaxation. A
reasonable agreement in $\beta_1^{ord}$ can be observed. Part of the
statistical error in our dynamic measurements is from the input of
the bulk exponents $\nu_s$ and $z_s$.

With $m_1(t)$ at hand, one may proceed to investigate the
time-dependent susceptibility, $\chi_{11}(t)$. In the case of the
ordinary transition of the $3d$ Ising model, however, $\gamma_{11}$
is negative. Therefore the $\chi_{11}$ is suppressed during the time
relaxation according to Eq. (\ref{e50}) and fluctuating around $0$
if nonequilirium preparation is an ordered initial state
$\chi_{11}(0)=0$. The power-law behavior in Eq.~(\ref{e50}) could
not be observed. Nevertheless, at the special transition, where
$\gamma_{11}$ is positive, the situation is different. The power-law
behavior of the surface susceptibility and cumulant shows up.

In Fig. \ref{f1a} and \ref{f1b}, the surface magnetization, surface
susceptibility and appropriate cumulant are displayed at the special
transition $J_s/J_b=r_{sp}$. A power-law behavior is observed for
all three observables. From the slope of the curve of the surface
magnetization, we measure $\beta_{1}^{sp}/\nu_s z_s=0.171(2)$, and
then obtain $\beta_{1}^{sp}=0.220(3)$ with $\nu_s$ and $z_s$ in
Table \ref{tbl1} as input. From the curve of the surface
susceptibility, we measure $\gamma_{11}^{sp}/\nu_s z_s=0.640(3)$,
and then calculate $\gamma_{11}^{sp}=0.823(4)$. From the scaling law
$\gamma_{11}/\nu_s=d-1-2\beta_1/\nu_s$, one derives
$\beta_{1}^{sp}=0.218(2)$, which is in good agreement with
$\beta_{1}^{sp}=0.220(3)$ estimated from the surface magnetization.
The scaling behaviors in Eqs.~(\ref{e30}) and (\ref{e50}) indeed
hold.

The remarkable feature of the cumulant {\it on the surface} is that
its scaling behavior in Eq.~(\ref{e60}) does not involve the
exponent $\beta_1$ of the surface magnetization. From the curve in
Fig. \ref{f1b}, we obtain $(d-1)/z_s=0.996(11)$, then calculate the
{\it bulk} dynamic critical exponent $z_s=2.01(2)$. This value of
$z_s$ is very close to $z_{3d}=2.04(1)$ measured in numerical
simulations in the bulk in Table~\ref{tbl1}, and it confirms that
the dynamic exponent on the surface is the same as that in the bulk.

\begin{figure}[t]
\psfig{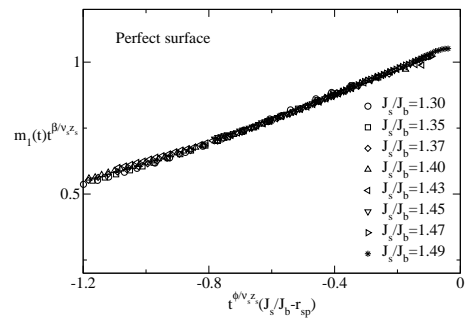} \caption{\label{f2}The
scaling plot of $m_1(t)$ according to Eq.~(\ref{neqcs}) around the
special transition $J_s/J_b=r_{sp}=1.5004$. The time window in this
plot is within $[10,1000]$. The lattice size is $L=80$ and $T=T_c$.}
\end{figure}

In order to describe the dynamic behavior of the surface
magnetization around $r_{sp}$, we need to introduce a crossover
scaling relation. To understand the scaling relation in
non-equilibrium states, we first recall the crossover scaling
relation in equilibrium. In equilibrium, $m_1(\tau)$ near the
special transition is described by a crossover scaling relation
\begin{equation}\label{eqcs}
m_1(\tau)\tau^{-\beta_{1}^{sp}}=M_{eq}(\tau^{-\phi}(J_s/J_b-r_{sp})),
\end{equation}
where $\tau=1-T/T_c$ is the reduced temperature, and $\phi$ is the
crossover exponent. From the crossover scaling relation of
$m_1(\tau)$ , one can determine the special transition point
$r_{sp}$ as well as $\beta_{1}^{sp}$ and $\phi$ \cite{lan90}.
Nevertheless, up to now it has not been studied whether there also
exists a corresponding crossover scaling relation in non-equilibrium
states. Here we will verify that such a dynamic crossover scaling
form indeed exists. For simplicity, we consider the case when
$J_s/J_b$ approaches the special transition from $r_{sp}^{-}$ and
the system is at the bulk critical temperature. Now the
non-equilibrium spatial correlation length $\xi(t) \sim t^{1/z}$
takes the place of the equilibrium spatial correlation length
$\tau^{-\nu}$. By substituting $t^{-1/\nu_s z_s}$ for $\tau$ into
Eq.~(\ref{eqcs}), we obtain
\begin{equation}\label{neqcs}
m_1(t)t^{\beta_{1}^{sp}/\nu_s z_s}=M_{neq}(t^{\phi/\nu_s
z_s}(J_s/J_b-r_{sp})).
\end{equation}
We have performed non-equilibrium simulations at $J_s/J_b=1.30$,
$1.35$, $1.37$, $1.40$, $1.43$, $1.45$, $1.47$, $1.49$, and made a
scaling plot according to Eq.~(\ref{neqcs}). This is demonstrated in
Fig. \ref{f2}. All curves of different $J_s/J_b$  collapse into a
single master curve, and it indicates that Eq.~(\ref{neqcs}) {\it
does} describe the crossover behavior during the dynamic relaxation.
The scaling plot in Fig.~\ref{f2} yields the exponents $\phi=0.52$
and $\beta_1^{sp}=0.220$, as well as the special transition point
$r_{sp}=1.50$. The crossover exponent $\phi$ is very close to the
mean-field value $0.5$ \cite{lan90}, and $\beta_1^{sp}$ and $r_{sp}$
are in agreement with the existing results from simulations in
equilibrium in Table \ref{tbl2} and in Ref. \cite{rug92}. Although
the precision of $r_{sp}$ and critical exponents obtained here are
not very high, it is still theoretically interesting. The dynamic
crossover scaling form in Eq.~(\ref{neqcs}) should be general, and
hold in various statistical systems.

\begin{figure}[t]
\psfig{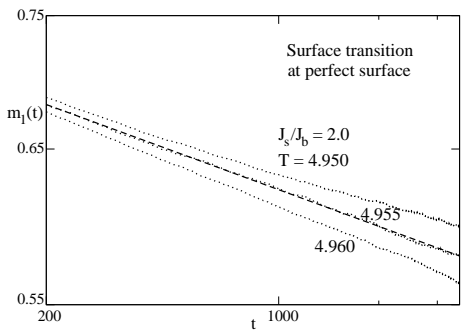}
\caption{\label{f3a}Determination of the surface transition
temperature $T_{s}$ for $J_s/J_b=2.0$. The dashed line is a
power-law fit to the curve of $T=4.955$. The lattice size is
$L=80$.}
\end{figure}

\begin{figure}[b]
\psfig{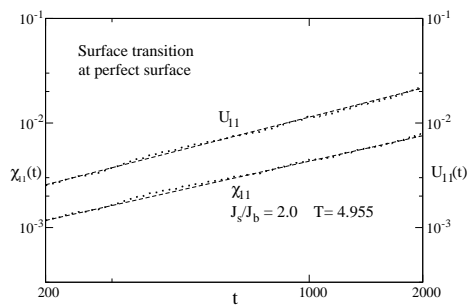}
\caption{\label{f3b}Dynamic relaxation of the surface susceptibility
and cumulant at the surface transition $J_s/J_b=2.0$ plotted on
double-log scale. The lattice size is $L=80$ and $T_s=4.955$. The
dashed lines are power-law fits to the curves.}
\end{figure}

To carry out the simulation at the surface transition, we fix
$J_s/J_b$ at $2.0$, well above $r_{sp}$. At the surface transition,
where the critical fluctuation is essentially two dimensional,
$\nu_{s}$ and $z_{s}$ in Eq.~(\ref{e30}) become $\nu_{2d}$ and
$z_{2d}$. Around the transition temperature, the surface
magnetization obeys a dynamic scaling form $<m_1(t)>\sim
t^{-\beta_1/\nu_s z_s}F(t^{1/\nu_s z_s}\tau)$ \cite{zhe98}. To
determine the surface transition temperature $T_s$, one may search
for a best-fitting power-law curve to the surface magnetization.
Then the corresponding temperature is identified as the transition
temperature $T_s$. We perform the simulations with three
temperatures around the transition temperature $T_s$, and measure
the surface magnetization. The results are displayed in Fig.
\ref{f3a}. Interpolating the surface magnetization to other
temperatures around these three temperatures, one finds the best
power-law behavior of the surface magnetization at $T_s=4.955$. The
corresponding slope of the curve gives $\beta_1/z_{s}=0.0570(10)$ at
$T_s=4.955$, and it is in agreement with the value in the $2d$ Ising
model \cite{zhe98}. Therefore we take $T_s=4.955$ as the surface
transition temperature, which is consistent with $T_s=4.9575(75)$
obtained with Monte Carlo simulations in equilibrium \cite{ple99}.

The time-dependent cumulant $U$ and susceptibility $\chi_{11}$ at
the surface transition are measured, and displayed in Fig.
\ref{f3b}. The slope of cumulant is $0.916(15)$, in a good agreement
with $2/z_{2d}=0.926(9)$ of the $2d$ Ising model \cite{zhe98}.
Consistence is also observed for the susceptibility where the slope
is $0.824(10)$, in comparison with $\gamma_{2d}/z_{2d}=0.810(8)$ in
the $2d$ Ising model. We thus confirm that the surface transition
belongs to the universality class of the $2d$ Ising model.
Meanwhile, $T_s=4.955$ is a good estimate of the surface transition
temperature.

\section{Short-time dynamics on a non-perfect surface}\label{sec4}

In this section we investigate the nonequilibrium critical dynamics
on a non-perfect surface, i.e. $J_l\neq J_s$. The static and dynamic
properties of a non-perfect surface are important and interesting,
because real surfaces are often rough, due to the impurity or
limitation of experimental conditions \cite{ple04}. Furthermore, the
advance in nano-science allows experimentalists to create other
structures on top of films artificially. We study the line defect on
a surface, and the procedure can be generalized to other extended
defects.

\begin{figure}[b]
\psfig{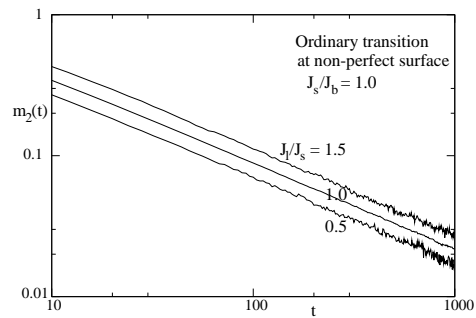}
\caption{\label{f4a}Dynamic relaxation of the line magnetization at
the ordinary transition on a non-perfect surface with various $J_l$
is plotted on a double-log scale. The slope of the curves is
independent of $J_l$.}
\end{figure}

We first consider the dynamic behavior of $m_2$ at the ordinary
transition. For convenience, we fix $J_s/J_b=1.0$. The profiles of
$m_2(t)$ with $J_l=0.5J_s$, $J_l=1.0J_s$ and $J_l=1.5J_s$ are
depicted in Fig. \ref{f4a}. All lines look parallel to each other,
and it indicates that they may belong to a same universality class.
By fitting these curves to the power law in Eq.~(\ref{e30}), we
estimate $\beta_2^{ord}=0.792(18)$, $0.786(6)$ and $0.797(33)$ with
$J_l=1.5J_s$, $J_l=1.0J_s$ and $J_l=0.5J_s$ respectively. These
values are consistent with each other and with $\beta_1^{ord}$ on
the perfect surface reported in the previous section. It confirms
that the defect in the ordinary transition is irrelevant, in term of
the renormalization group argument. This conclusion echoes that in
Ref. \cite{ple98b}, where the impact of random bonds on the surface
is investigated {\it in equilibrium}. According to the generalized
Harris criterion \cite{die90}, defects with random bonds or diluted
bonds on a surface are irrelevant. The short-time dynamic approach
shows its merits in identifying the universal behavior of the
surface magnetization \cite{ple98b,ple98,ple99,ple00}. Here we note
that the line magnetization is one-dimensional, and therefore
somewhat more fluctuating than the surface magnetization.

\begin{figure}[t]
\psfig{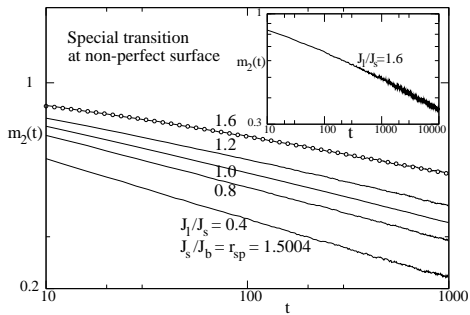}
\caption{\label{f4b}Dynamic relaxation of the line magnetization at
the special transition $J_s/J_b=r_{sp}=1.5004$ on a non-perfect
surface with various $J_l$ is plotted on a double-log scale. The
open circles are a fit to Eq.~(\ref{e70}) with a correction to
scaling. The inset displays the line magnetization at $J_l=1.6J_s$
but with a longer simulation time. The lattice size is $L=128$ and
$T=T_c$. The slope of the curves is dependent on $J_l$ even after
taking the correction to scaling into account.}
\end{figure}

\begin{figure}[b]
\psfig{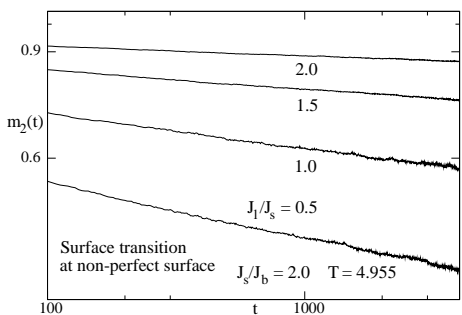}
\caption{\label{f5}Dynamic relaxation of the line magnetization with
various $J_l$ at the surface transition is plotted on a double-log
scale. The lattice size is $L=80$ and $T=T_s$. The slope of the
curves is dependent on $J_l$, and given in Table \ref{tbl3}.}
\end{figure}

Now we turn to the special transition. We perform simulations with
various $J_l$ at the special transition, and the line magnetization
is presented in Fig. \ref{f4b}. From the slopes of the curves, one
measures the exponent $\beta_2^{sp}/\nu_s z_s$, and then calculates
$\beta_{2}^{sp}=0.260(4)$, $0.230(3)$, $0.219(5)$, $0.204(6)$ and
$0.162(3)$ for $J_l=0.4J_s$, $J_l=0.8J_s$, $J_l=1.0J_s$,
$J_l=1.2J_s$ and $J_l=1.6J_s$ respectively, with $\nu_s$ and $z_s$
taken as input from Table \ref{tbl1}. Obviously $\beta_{2}^{sp}$
changes continuously with $J_l$.

\begin{figure}[t]
\psfig{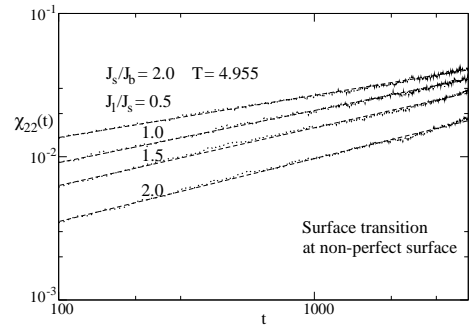}
\caption{\label{f6a}Dynamic relaxation of the line susceptibility
with various $J_l$ at the surface transition plotted on a double-log
scale. The dashed lines are power-law fits.  The slope of the curves
is dependent on $J_l$, and given in Table \ref{tbl3}.}
\end{figure}

Since there exists certain deviation from a power law in shorter
times for the curves with a larger ratio $J_l/J_s$ in Fig.
\ref{f4b}, one may wonder whether the small variation in
$\beta_{2}^{sp}$ may stem from the correction to scaling induced by
the defect line. Therefore, a careful analysis of the correction to
scaling is necessary in this case. Assuming a power-law correction
to scaling, $m_2(t)$ should evolve according to
\begin{equation}
m_2(t)=at^{-\beta_{2}^{sp}/\nu_s z_s}(1-bt^{-c}) .\label{e70}
\end{equation}
As shown in Fig. \ref{f4b}, such an ansatz fits the numerical data
very well, and yields $\beta_{2}^{sp}=0.258(1)$, $0.235(7)$,
$0.228(3)$, $0.214(6)$ and $0.171(3)$ for $J_l=0.4J_s$,
$J_l=0.8J_s$, $J_l=1.0J_s$, $J_l=1.2J_s$ and $J_l=1.6J_s$
respectively. For $J_l=1.6J_s$, we extend our simulations up to a
maximum time $t=10000$MCS to gain more confidence on our results.
Still $\beta_{2}^{sp}$ varies continuously with $J_l$, and the
strong universality is violated. This is different from the case on
a random surface, where the generalized Harris criterion states that
the enhancement of the short-range randomness on the surface is
irrelevant at the surface transition in the $3d$ Ising model
\cite{die90}. Our result is, however, not surprising, for the defect
line is not a short-range randomness \cite{ple04}, but an extended
one. As the short-range random surface is close to being relevant
\cite{ple04}, it is not surprising that the defect line modifies the
surface universality class. This can also be understood. The
reduction of the coupling in the defect line is somewhat like
turning the local surface from the special transition to the
ordinary one, and therefore gives rise to a large value of the
critical exponent $\beta_{2}^{sp}$.

\begin{table*}[t]\centering
\caption{Comparison between the numerical simulations of surface
transition with a non-perfect surface and the theory of the
two-dimensional Ising model with a defect line. $\nu_s=\nu_{2d}=1$
and $z_{s}=z_{2d}=2.16(2)$ have been taken as input \cite{zhe98}.}
\begin{tabular}{c | c  c | c  c | c  c }\hline\hline
 & line magnetization& & susceptibility & & cumulant&\\
\hline
exponent & $\beta_2/z_{s}$& &$(1-2\beta_2)/z_{s}$& & $1/z_{s}$&\\
\hline  & Simulation & Theory & Simulation & Theory & Simulation &
Theory\\
\hline
$J_l=0.5J_s$ & $0.0923(36)$ & $0.0936(9)$ & $0.282(4)$ & $0.276(3)$ & $0.462(2)$ & $0.463(4)$\\
$J_l=1.0J_s$ & $0.0570(10)$ & $0.0579(5)$ & $0.356(5)$ & $0.347(3)$ & $0.475(2)$ & $0.463(4)$\\
$J_l=1.5J_s$ & $0.0301(24)$ & $0.0307(3)$ & $0.405(4)$ & $0.402(4)$ & $0.468(2)$ & $0.463(4)$\\
$J_l=2.0J_s$ & $0.0149(12)$ & $0.0145(1)$ & $0.428(9)$ & $0.434(4)$ & $0.459(9)$ & $0.463(4)$\\
 \hline \hline
\end{tabular}
\label{tbl3}
\end{table*}

To investigate the impact of the line defect at the surface
transition, we fix $J_s/J_b=2.0$. We measure the time evolution of
the line magnetization at its transition temperature $T_s=4.955$
with $J_l=0.5J_s$, $J_l=1.0J_s$, $J_l=1.5J_s$ and $J_l=2.0J_s$. In
Fig. \ref{f5}, one observes that after a microscopic time $t_{mic}
\sim 100$MCS, the power-law behavior emerges. However, the exponent
$\beta_2$ is $J_l$-dependent, and the strong universality is
violated. This is similar to the case in Ref. \cite{ple99,ple00},
where a non-universal behavior of the edge and corner magnetization
has been found at the surface transition.

\begin{figure}[b]
\psfig{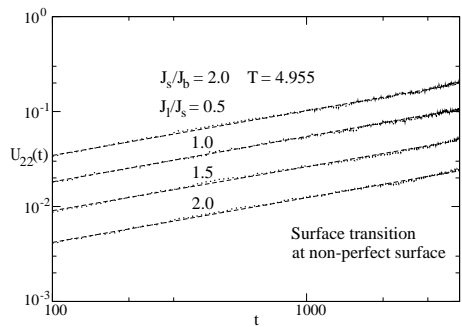}
\caption{\label{f6b}Dynamic relaxation of the cumulant with various
$J_l$ at the surface transition plotted on a double-log scale. The
dashed lines are power-law fits. The slope of the curves is
independent of $J_l$, and given in Table \ref{tbl3}.}
\end{figure}

Since the surface transition is essentially two-dimensional, one may
relate this non-perfect surface to the $2d$ Ising model with a
defect line {\it without the presence of bulk}. The violation of the
strong universality of the $2d$ Ising model with a line or a ladder
defect is rigorously proved by Bariev \cite{bar79}. For the line
defect, exact calculations show that
\begin{equation}\label{exact1}
\beta_2=\frac{2}{\pi^2}\arctan^2(\kappa_l),
\end{equation}
with
\begin{equation}\label{exact2}
\kappa_l=\exp(-2(J_l-J)/k_BT_{c}).
\end{equation}
The critical exponent $\beta_2$ reduces monotonically, when the
defect coupling $J_l$ is enhanced. We measure the exponent $\beta_2$
and compare it with the exact values obtained from Eqs.
(\ref{exact1}) and (\ref{exact2}). The results are summarized in
Table \ref{tbl3}. One finds a good agreement between simulations and
exact results. A similar behavior of the edge magnetization, which
can be viewed as a line defect at the surface transition, is also
observed in Ref. \cite{ple99}. Our results support that at the
surface transition, the critical exponent $\beta_2$ will change in
the presence of a small perturbation.

Finally, the susceptibility $\chi_{22}(t)$ and cumulant $U_{22}(t)$
of the line magnetization, which are similarly defined as those of
the surface magnetization, are also measured. The results are
plotted in Fig. \ref{f6a} and \ref{f6b}. Simple scaling analysis
shows that $\chi_{22}(t)\sim t^{(d-2-2\beta_2/\nu_s)/z_s}$ and
$U_{22}(t)\sim t^{(d-2)/z_s}$. The estimated exponents are also
compiled in Table \ref{tbl3}, and a good consistency with the theory
can be spotted.

\section{Conclusion}\label{sec5}

With Monte Carlo simulations, we have studied the dynamic relaxation
on a perfect and non-perfect surface in the $3d$ Ising model,
starting from an ordered initial state. On the perfect surface, the
dynamic behavior of the surface magnetization, susceptibility and
appropriate cumulant is carefully analyzed at the ordinary, special
and surface transition. The universal dynamic scaling behavior is
revealed, and the static exponent $\beta_1$ of the surface
magnetization, the static exponent $\gamma_{11}$ of the surface
susceptibility and the dynamic exponent $z_s$ are estimated. All the
results for $\beta_1$ are compiled in Table \ref{tbl2}. Since the
exponents $\nu_s$ and $z_s$ can be identified as those at bulk, it
is convenient to study different phase transitions from the
non-equilibrium dynamic relaxation. Especially, the dynamic
crossover scaling form in Eq.~(\ref{neqcs}) is interesting. Because
of the existence of new scaling variable $J_s/J_b$, the
nonequilibrium relaxation of magnetization at the critical
temperature may not obey a power law, which is quite different from
the general systems investigated so far where a power law behavior
was always expected. This unusual nonequilibrium behavior is a
consequence of the presence of geometric surface.

On the non-perfect surface, i.e., with a defect line in the surface,
the universality class of the ordinary transition remains the same
as that at the perfect surface. On the other hand, for the special
and surface transitions, the critical exponent $\beta_2$ of the line
magnetization varies with the coupling $J_l$ strength of the defect
line. The susceptibility and appropriate cumulant of the line
magnetization also exhibit the dynamic scaling behavior and yield
the static exponent $\gamma_{22}$ and the dynamic exponent $z_s$.
The short-time dynamic approach is efficient in understanding the
surface critical phenomena.

{\bf Acknowledgements:} This work was supported in part by NNSF
(China) under Grant No. 10325520. The authors would like to thank M.
Pleimling for helpful discussions. One of the authors (SZL) would
like to thank L. Y. Wang for critical reading this manuscript. The
computations are partially carried out in Shanghai Supercomputer
Center.


\begin{thebibliography}{45}
\expandafter\ifx\csname
natexlab\endcsname\relax\def\natexlab#1{#1}\fi
\expandafter\ifx\csname bibnamefont\endcsname\relax
  \def\bibnamefont#1{#1}\fi
\expandafter\ifx\csname bibfnamefont\endcsname\relax
  \def\bibfnamefont#1{#1}\fi
\expandafter\ifx\csname citenamefont\endcsname\relax
  \def\citenamefont#1{#1}\fi
\expandafter\ifx\csname url\endcsname\relax
  \def\url#1{\texttt{#1}}\fi
\expandafter\ifx\csname urlprefix\endcsname\relax\def\urlprefix{URL
}\fi \providecommand{\bibinfo}[2]{#2}
\providecommand{\eprint}[2][]{\url{#2}}

\bibitem[{\citenamefont{{M.A. Torija, A.P. Li, X.C. Guan, E.W. Plummer, and J.
  Shen}}(2005)}]{tor05}
\bibinfo{author}{\bibnamefont{{M.A. Torija, A.P. Li, X.C. Guan, E.W. Plummer,
  and J. Shen}}}, \bibinfo{journal}{Phys. Rev. Lett.}
  \textbf{\bibinfo{volume}{95}}, \bibinfo{pages}{257203}
  (\bibinfo{year}{2005}).

\bibitem[{\citenamefont{{K. Binder}}(1987)}]{bin87}
\bibinfo{author}{\bibnamefont{{K. Binder}}}, \emph{\bibinfo{title}{in Phase
  Transition and Critical Phenomena}} (\bibinfo{publisher}{Academic Press,
  London}, \bibinfo{year}{1987}), \bibinfo{note}{vol. 8, p.1, and references
  therein.}

\bibitem[{\citenamefont{{H.W. Diehl}}(1987)}]{die87}
\bibinfo{author}{\bibnamefont{{H.W. Diehl}}}, \emph{\bibinfo{title}{in Phase
  Transition and Critical Phenomena}} (\bibinfo{publisher}{Academic Press,
  London}, \bibinfo{year}{1987}), \bibinfo{note}{vol. 10, p.76, and references
  therein.}

\bibitem[{\citenamefont{{H.W. Diehl}}(1997)}]{die97}
\bibinfo{author}{\bibnamefont{{H.W. Diehl}}}, \bibinfo{journal}{Int. J. Mod.
  Phys.} \textbf{\bibinfo{volume}{B11}}, \bibinfo{pages}{3503}
  (\bibinfo{year}{1997}).

\bibitem[{\citenamefont{{M. Pleimling}}(2004)}]{ple04}
\bibinfo{author}{\bibnamefont{{M. Pleimling}}}, \bibinfo{journal}{J. Phys.}
  \textbf{\bibinfo{volume}{A37}}, \bibinfo{pages}{R79} (\bibinfo{year}{2004}).

\bibitem[{\citenamefont{{K. Binder and P.C. Hohenberg}}(1972)}]{bin72}
\bibinfo{author}{\bibnamefont{{K. Binder and P.C. Hohenberg}}},
  \bibinfo{journal}{Phys. Rev.} \textbf{\bibinfo{volume}{B6}},
  \bibinfo{pages}{3461} (\bibinfo{year}{1972}).

\bibitem[{\citenamefont{{K. Binder and D.P. Landau}}(1984)}]{bin84}
\bibinfo{author}{\bibnamefont{{K. Binder and D.P. Landau}}},
  \bibinfo{journal}{Phys. Rev. Lett.} \textbf{\bibinfo{volume}{52}},
  \bibinfo{pages}{318} (\bibinfo{year}{1984}).

\bibitem[{\citenamefont{{D.P. Landau and K. Binder}}(1990)}]{lan90}
\bibinfo{author}{\bibnamefont{{D.P. Landau and K. Binder}}},
  \bibinfo{journal}{Phys. Rev.} \textbf{\bibinfo{volume}{B41}},
  \bibinfo{pages}{4633} (\bibinfo{year}{1990}).

\bibitem[{\citenamefont{{C. Ruge and F. Wagner}}(1995)}]{rug95}
\bibinfo{author}{\bibnamefont{{C. Ruge and F. Wagner}}},
  \bibinfo{journal}{Phys. Rev.} \textbf{\bibinfo{volume}{B52}},
  \bibinfo{pages}{4209} (\bibinfo{year}{1995}).

\bibitem[{\citenamefont{{Y. Deng, H.W.J. Bl\"ote, and M.P.
  Nightingale}}(2005)}]{den05}
\bibinfo{author}{\bibnamefont{{Y. Deng, H.W.J. Bl\"ote, and M.P.
  Nightingale}}}, \bibinfo{journal}{Phys. Rev.} \textbf{\bibinfo{volume}{E72}},
  \bibinfo{pages}{016128} (\bibinfo{year}{2005}).

\bibitem[{\citenamefont{{S. Dietrich and H.W. Diehl}}(1983)}]{die83}
\bibinfo{author}{\bibnamefont{{S. Dietrich and H.W. Diehl}}},
  \bibinfo{journal}{Z. Phys.} \textbf{\bibinfo{volume}{B51}},
  \bibinfo{pages}{343} (\bibinfo{year}{1983}).

\bibitem[{\citenamefont{{M. Kikuchi and Y. Okabe}}(1985)}]{kik85}
\bibinfo{author}{\bibnamefont{{M. Kikuchi and Y. Okabe}}},
  \bibinfo{journal}{Phys. Rev. Lett.} \textbf{\bibinfo{volume}{55}},
  \bibinfo{pages}{1220} (\bibinfo{year}{1985}).

\bibitem[{\citenamefont{{H.W. Diehl}}(1994)}]{die94}
\bibinfo{author}{\bibnamefont{{H.W. Diehl}}}, \bibinfo{journal}{Phys. Rev.}
  \textbf{\bibinfo{volume}{B49}}, \bibinfo{pages}{2846} (\bibinfo{year}{1994}).

\bibitem[{\citenamefont{{H.K. Janssen, B. Schaub and B.
  Schmittmann}}(1989)}]{jan89}
\bibinfo{author}{\bibnamefont{{H.K. Janssen, B. Schaub and B. Schmittmann}}},
  \bibinfo{journal}{Z. Phys.} \textbf{\bibinfo{volume}{{B 73}}},
  \bibinfo{pages}{539} (\bibinfo{year}{1989}).

\bibitem[{\citenamefont{Huse}(1989)}]{hus89}
\bibinfo{author}{\bibfnamefont{D.}~\bibnamefont{Huse}}, \bibinfo{journal}{Phys.
  Rev.} \textbf{\bibinfo{volume}{{B 40}}}, \bibinfo{pages}{304}
  (\bibinfo{year}{1989}).

\bibitem[{\citenamefont{Zheng}(1998)}]{zhe98}
\bibinfo{author}{\bibfnamefont{B.}~\bibnamefont{Zheng}}, \bibinfo{journal}{Int.
  J. Mod. Phys.} \textbf{\bibinfo{volume}{B12}}, \bibinfo{pages}{1419}
  (\bibinfo{year}{1998}), \bibinfo{note}{review article}.

\bibitem[{\citenamefont{{B. Zheng, M. Schulz, and S. Trimper}}(1999)}]{zhe99}
\bibinfo{author}{\bibnamefont{{B. Zheng, M. Schulz, and S. Trimper}}},
  \bibinfo{journal}{Phys. Rev. Lett.} \textbf{\bibinfo{volume}{{82}}},
  \bibinfo{pages}{1891} (\bibinfo{year}{1999}).

\bibitem[{\citenamefont{{C. Godr\'eche and J.M. Luck}}(2002)}]{god02}
\bibinfo{author}{\bibnamefont{{C. Godr\'eche and J.M. Luck}}},
  \bibinfo{journal}{J. Phys.:Condens. Matter} \textbf{\bibinfo{volume}{14}},
  \bibinfo{pages}{1589} (\bibinfo{year}{2002}).

\bibitem[{\citenamefont{{P. Calabrese and A. Gambassi}}(2005)}]{cal05}
\bibinfo{author}{\bibnamefont{{P. Calabrese and A. Gambassi}}},
  \bibinfo{journal}{J. Phys.} \textbf{\bibinfo{volume}{A38}},
  \bibinfo{pages}{R133} (\bibinfo{year}{2005}).

\bibitem[{\citenamefont{{Z.B. Li, L. {Sch\"ulke}, and B. Zheng}}(1995)}]{li95}
\bibinfo{author}{\bibnamefont{{Z.B. Li, L. {Sch\"ulke}, and B. Zheng}}},
  \bibinfo{journal}{Phys. Rev. Lett.} \textbf{\bibinfo{volume}{{74}}},
  \bibinfo{pages}{3396} (\bibinfo{year}{1995}).

\bibitem[{\citenamefont{{H.J. Luo, L. Sch\"ulke, and B. Zheng}}(1998)}]{luo98}
\bibinfo{author}{\bibnamefont{{H.J. Luo, L. Sch\"ulke, and B. Zheng}}},
  \bibinfo{journal}{Phys. Rev. Lett.} \textbf{\bibinfo{volume}{{81}}},
  \bibinfo{pages}{180} (\bibinfo{year}{1998}).

\bibitem[{\citenamefont{{L. Sch\"ulke and B. Zheng}}(2000)}]{sch00}
\bibinfo{author}{\bibnamefont{{L. Sch\"ulke and B. Zheng}}},
  \bibinfo{journal}{Phys. Rev.} \textbf{\bibinfo{volume}{{E62}}},
  \bibinfo{pages}{7482} (\bibinfo{year}{2000}).

\bibitem[{\citenamefont{{B. Zheng, F. Ren, and H. Ren}}(2003)}]{zhe03a}
\bibinfo{author}{\bibnamefont{{B. Zheng, F. Ren, and H. Ren}}},
  \bibinfo{journal}{Phys. Rev.} \textbf{\bibinfo{volume}{{E68}}},
  \bibinfo{pages}{046120} (\bibinfo{year}{2003}).

\bibitem[{\citenamefont{{J.Q. Yin, B. Zheng, and S. Trimper}}(2004)}]{yin04}
\bibinfo{author}{\bibnamefont{{J.Q. Yin, B. Zheng, and S. Trimper}}},
  \bibinfo{journal}{Phys. Rev.} \textbf{\bibinfo{volume}{E70}},
  \bibinfo{pages}{056134} (\bibinfo{year}{2004}).

\bibitem[{\citenamefont{{J.Q. Yin, B. Zheng, and S. Trimper}}(2005)}]{yin05}
\bibinfo{author}{\bibnamefont{{J.Q. Yin, B. Zheng, and S. Trimper}}},
  \bibinfo{journal}{Phys. Rev.} \textbf{\bibinfo{volume}{E72}},
  \bibinfo{pages}{036122} (\bibinfo{year}{2005}).

\bibitem[{\citenamefont{{A.A. Fedorenko and S. Trimper}}(2006)}]{fed06}
\bibinfo{author}{\bibnamefont{{A.A. Fedorenko and S. Trimper}}},
  \bibinfo{journal}{Europhys. Lett.} \textbf{\bibinfo{volume}{74}},
  \bibinfo{pages}{89} (\bibinfo{year}{2006}).

\bibitem[{\citenamefont{Ritschel and Czerner}(1995)}]{rit95}
\bibinfo{author}{\bibfnamefont{U.}~\bibnamefont{Ritschel}} \bibnamefont{and}
  \bibinfo{author}{\bibfnamefont{P.}~\bibnamefont{Czerner}},
  \bibinfo{journal}{Phys. Rev. Lett.} \textbf{\bibinfo{volume}{{75}}},
  \bibinfo{pages}{3882} (\bibinfo{year}{1995}).

\bibitem[{\citenamefont{{M. Pleimling and F. Igl\'{o}i}}(2004)}]{ple04b}
\bibinfo{author}{\bibnamefont{{M. Pleimling and F. Igl\'{o}i}}},
  \bibinfo{journal}{Phys. Rev. Lett.} \textbf{\bibinfo{volume}{92}},
  \bibinfo{pages}{145701} (\bibinfo{year}{2004}).

\bibitem[{\citenamefont{{M. Pleimling and F. Igl\'{o}i}}(2005)}]{ple05}
\bibinfo{author}{\bibnamefont{{M. Pleimling and F. Igl\'{o}i}}},
  \bibinfo{journal}{Phys. Rev.} \textbf{\bibinfo{volume}{B71}},
  \bibinfo{pages}{094424} (\bibinfo{year}{2005}).

\bibitem[{\citenamefont{{H.J. Luo, L. Sch\"ulke, B. Zheng}}(2001)}]{luo01}
\bibinfo{author}{\bibnamefont{{H.J. Luo, L. Sch\"ulke, B. Zheng}}},
  \bibinfo{journal}{Phys. Rev.} \textbf{\bibinfo{volume}{{E64}}},
  \bibinfo{pages}{036123} (\bibinfo{year}{2001}).

\bibitem[{\citenamefont{{C. Ruge, S. Dunkelmann, and F. Wagner}}(1992)}]{rug92}
\bibinfo{author}{\bibnamefont{{C. Ruge, S. Dunkelmann, and F. Wagner}}},
  \bibinfo{journal}{Phys. Rev. Lett.} \textbf{\bibinfo{volume}{69}},
  \bibinfo{pages}{2465} (\bibinfo{year}{1992}).

\bibitem[{\citenamefont{{M. Pleimling and W.
  Selke}}(1998{\natexlab{a}})}]{ple98b}
\bibinfo{author}{\bibnamefont{{M. Pleimling and W. Selke}}},
  \bibinfo{journal}{Eur. Phys. J.} \textbf{\bibinfo{volume}{B1}},
  \bibinfo{pages}{385} (\bibinfo{year}{1998}{\natexlab{a}}).

\bibitem[{\citenamefont{{H.W. Diehl and A. N\"usser}}(1990)}]{die90}
\bibinfo{author}{\bibnamefont{{H.W. Diehl and A. N\"usser}}},
  \bibinfo{journal}{Z. Phys.} \textbf{\bibinfo{volume}{B79}},
  \bibinfo{pages}{69} (\bibinfo{year}{1990}).

\bibitem[{\citenamefont{{H.W. Diehl}}(1998)}]{die98a}
\bibinfo{author}{\bibnamefont{{H.W. Diehl}}}, \bibinfo{journal}{Eur. Phys. J.}
  \textbf{\bibinfo{volume}{B1}}, \bibinfo{pages}{401} (\bibinfo{year}{1998}).

\bibitem[{\citenamefont{{M.E. Fisher and A.E. Ferdinand}}(1967)}]{fis67}
\bibinfo{author}{\bibnamefont{{M.E. Fisher and A.E. Ferdinand}}},
  \bibinfo{journal}{Phys. Rev. Lett.} \textbf{\bibinfo{volume}{19}},
  \bibinfo{pages}{169} (\bibinfo{year}{1967}).

\bibitem[{\citenamefont{{Y. Deng and H.W.J. Bl\"ote}}(2003)}]{den03}
\bibinfo{author}{\bibnamefont{{Y. Deng and H.W.J. Bl\"ote}}},
  \bibinfo{journal}{Phys. Rev.} \textbf{\bibinfo{volume}{E68}},
  \bibinfo{pages}{036125} (\bibinfo{year}{2003}).

\bibitem[{\citenamefont{{A.M. Ferrenberg and D.P. Landau}}(1991)}]{fer91}
\bibinfo{author}{\bibnamefont{{A.M. Ferrenberg and D.P. Landau}}},
  \bibinfo{journal}{Phys. Rev.} \textbf{\bibinfo{volume}{B44}},
  \bibinfo{pages}{5081} (\bibinfo{year}{1991}).

\bibitem[{\citenamefont{{A. Jaster, J. Mainville, L. Sch\"ulke, and B.
  Zheng}}(1999)}]{jas99}
\bibinfo{author}{\bibnamefont{{A. Jaster, J. Mainville, L. Sch\"ulke, and B.
  Zheng}}}, \bibinfo{journal}{J. Phys.} \textbf{\bibinfo{volume}{{A32}}},
  \bibinfo{pages}{1395} (\bibinfo{year}{1999}).

\bibitem[{\citenamefont{{B. Zheng}}(1996)}]{zhe96}
\bibinfo{author}{\bibnamefont{{B. Zheng}}}, \bibinfo{journal}{Phys. Rev. Lett.}
  \textbf{\bibinfo{volume}{{77}}}, \bibinfo{pages}{679} (\bibinfo{year}{1996}).

\bibitem[{\citenamefont{{M. Pleimling and W.
  Selke}}(1998{\natexlab{b}})}]{ple98}
\bibinfo{author}{\bibnamefont{{M. Pleimling and W. Selke}}},
  \bibinfo{journal}{Eur. Phys. J.} \textbf{\bibinfo{volume}{B5}},
  \bibinfo{pages}{805} (\bibinfo{year}{1998}{\natexlab{b}}).

\bibitem[{\citenamefont{{C. Ruge, A. Dunkelmann, F. Wagner and J.
  Wulf}}(1993)}]{rug93}
\bibinfo{author}{\bibnamefont{{C. Ruge, A. Dunkelmann, F. Wagner and J.
  Wulf}}}, \bibinfo{journal}{J. Stat. Phys.} \textbf{\bibinfo{volume}{110}},
  \bibinfo{pages}{1411} (\bibinfo{year}{1993}).

\bibitem[{\citenamefont{{H.W. Diehl and M. Shpot}}(1998)}]{die98}
\bibinfo{author}{\bibnamefont{{H.W. Diehl and M. Shpot}}},
  \bibinfo{journal}{Nucl. Phys.} \textbf{\bibinfo{volume}{B528}},
  \bibinfo{pages}{595} (\bibinfo{year}{1998}).

\bibitem[{\citenamefont{{M. Pleimling and W. Selke}}(1999)}]{ple99}
\bibinfo{author}{\bibnamefont{{M. Pleimling and W. Selke}}},
  \bibinfo{journal}{Phys. Rev.} \textbf{\bibinfo{volume}{B59}},
  \bibinfo{pages}{65} (\bibinfo{year}{1999}).

\bibitem[{\citenamefont{{M. Pleimling and W. Selke}}(2000)}]{ple00}
\bibinfo{author}{\bibnamefont{{M. Pleimling and W. Selke}}},
  \bibinfo{journal}{Phys. Rev.} \textbf{\bibinfo{volume}{E61}},
  \bibinfo{pages}{933} (\bibinfo{year}{2000}).

\bibitem[{\citenamefont{{R. Z. Bariev}}(1979)}]{bar79}
\bibinfo{author}{\bibnamefont{{R. Z. Bariev}}}, \bibinfo{journal}{Sov. Phys.
  JETP} \textbf{\bibinfo{volume}{50}}, \bibinfo{pages}{613}
  (\bibinfo{year}{1979}).

\end{thebibliography}

\end{document}